# Functionalization of Silicon Surface by Thiadiazole Molecule : a DFT Study

Çağıl KADEROĞLU[1,*]

[1] Department of Physics Engineering, Ankara University, Ankara, Turkey

*Corresponding author: cagil.kaderoglu@eng.ankara.edu.tr

**Abstract**

The first principles density functional theory (DFT) calculations have been used to investigate the atomic and electronic properties of thiadiazole adsorption on the Si(001) surface. A (2×2) reconstructed clean substrate surface has been chosen to give the molecule sufficient space to relax into its most favorable position. A total of eighteen bonding model including bridge-type bonding, and [2+2]/[4+2] cycloaddition mechanisms for four structural isomers of thiadiazole molecule have been discussed in these calculations. The most stable bonding configuration for each of the four isomers on the clean silicon surface has been determined by performing total energy calculations. Electronic structures of the stable surfaces have been investigated by plotting the total density of states (TDOS) and energy band diagrams. Charge densities have been plotted to determine the origin of the surface states appeared in the fundamental band gap of silicon. Finally, partial density of states (PDOS) have been plotted to see the contributions from s- and p-orbitals.

# 1. Introduction

Thiadiazoles within the family of azole compounds are five-membered heterocyclic aromatic molecules whose chemical formula is given as $C_2H_2N_2S$. This molecule has four structural isomers so-called 1,2,3-, 1,2,4-, 1,2,5- and 1,3,4-Thiadiazole according to the position of sulphur and nitrogen atoms in the ring, as shown in Fig. 1 [1]. The aromaticity of these isomers, which is an important parameter for determining the chemical reactivity and physical properties of the molecules and thus their potential applications in technology, can be ordered as 1,2,5-Thiadiazole > 1,2,3-Thiadiazole > 1,2,4-Thiadiazole > 1,3,4-Thiadiazole [2,3].

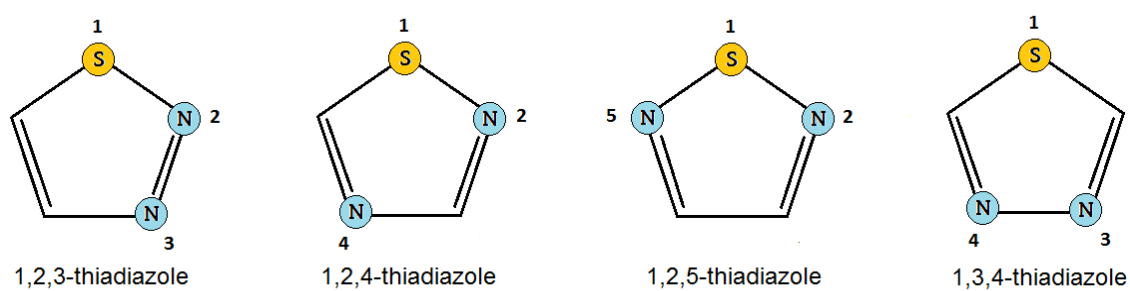

**Figure 1.** Structural isomers of thiadiazole

Since five- and six-membered heterocycles such as thiophenes, thiazoles and pyrazines have played an important role in technological research, monocyclic thiadiazole, which has similar electronic properties to these chemicals, has also started to be under the spotlight [4]. The thiadiazole ring acts as a scaffold between different molecule groups due to its electron-deficient nature. So, derivatives of thiadiazoles play a major role in advanced applications ranging from pharmaceutical chemistry to materials science [5-9]. Probably the most studied thiadiazole derivatives are the 1,3,4- and 1,2,4- ones which have applications in metal-ion-selective sensors and lithium rechargeable batteries in addition to their wide spectrum of biological activity [6,9-11]. Other technological uses of 1,3,4-Thiadiazole derivatives can be listed as corrosion and oxidation inhibitors, lubricant additives, optically active liquid crystals and optoelectronic materials [12-17]. 1,2,3-Thiadiazole derivatives are not only widely used in medical applications such as other isomers, but also in photolithography and agriculture [18-20]. Investigations on 1,2,5-Thiadiazole derivatives have focused on its potential applications in organic semiconducting technology and in environmental studies such as constructing high-performance $CO_2$ adsorbents [21-23].

In order to meet the growing needs of developing technology, new materials whose electronic, optical and magnetic properties can be adjusted according to the application field are constantly being investigated. An important part of these investigations has focused on the modification of known inorganic semiconductor surfaces with organic molecules [24-31]. Silicon is the most known of these inorganic semiconductors. It is one of the cornerstones of the electronics industry. Its different oriented surfaces have become more prioritized in terms of technological applications, because of their different physical and chemical properties than the bulk itself. Such that, these surfaces can provide a wide range of precisely tailored properties due to their ability to be functionalized by an atom or molecule group.

Adsorption of some five- and six-membered heterocyclic molecules on silicon surface were investigated in the past [24, 32-34]. However, in the literature search, no theoretical or experimental information was found about the adsorption of the thiadiazoles. Considering the important role of silicon and thiadiazoles in technology, it may be useful for various applications to know about the properties of the new structure obtained by combining these two. For the above-mentioned reason, in this study, the effects of thiadiazole adsorption on the atomic and electronic properties of the Si(001) surface have been investigated by DFT approaches to create a prediction for the future experimental studies.

## 2. Materials and Methods

All the calculations in this study were carried out by using Vienna Ab-initio Simulation Package (VASP), which is a plane-wave pseudopotential approach implemented DFT code [35-39]. The electron-ion interaction was described by projector augmented wave (PAW) approach with a cut-off energy of 500eV [40]. Perdew-Burke-Ernzerhof (PBE) functional with generalized gradient approximation (GGA) was used for the exchange-correlation terms in the electron–electron interaction [41,42]. To find the optimal k-points, convergence tests for a (1x1x1) silicon unit cell were performed using different Monkhorst–Pack k-point meshes [43] ranging from (3×3×3) to (10×10×10), and it was seen that the total energy value of the simulation cell did not change after (8×8×8). Based on this, the Brillouin zone integration of the (2×2) substrate cell was done using a (4×4×1) k-point scheme. The conjugate-gradient algorithm was used to optimize the structures [44]. The convergence criteria for electronic minimization is $10^{-4}$eV while it is $10^{-3}$eV for the ionic minimization. Gaussian smearing method with the smearing width of 0.2 eV was used. The atomic structure and charge density visuals were created by using VESTA program [45].

## 3. Results and Discussion

### 3.1 Substrate Selection:  Clean Si(001)-(2×2) Surface

In previous experimental and theoretical studies, it has been shown that the Si(001) surface reconstructs c(4x2) and p(2×2) at low temperatures, and it transforms into a (2x1) reconstruction at room temperature [46-50]. But according to another theoretical study in the literature, (2×2) reconstruction with alternating dimer (semi-antiphase) is energetically more stable than (2x1) surface by approximately 0.24 eV/ dimer [51]. By also considering the earlier papers based on this knowledge [24,52], it has been decided to use Si(001)-(2×2) surface as the substrate for thiadiazole adsorption. Therefore, this ring molecule has sufficient space to relax into the most favorable position that minimizes the total energy of the whole structure. The substrate has been modelled as a supercell consisting of 18Å vacuum and silicon layers in 14Å thickness with a lattice parameter of 5.46Å. This lattice constant is compatible with the literature, since the experimental lattice parameter of silicon is 5.431Å [53] while its DFT calculated value varies from 5.37Å to 5.47Å according to the selection of exchange potential [24,47,48,54,55]. Each Si atom in the back surface of the bottom layer was saturated with two hydrogen atoms to prevent undesired interactions.

During relaxation, the bottom 5 layers of 11 layers of silicon were fixed to their bulk positions. All the remaining substrate atoms, adsorbate atoms and saturating H atoms were allowed to relax. The dimer length on the silicon surface after relaxation is 2.33Å with an approximate tilt angle of 18.8°. In the fundamental gap of projected bulk bands, four surface states originated from the dimer components of the Si(001)-(2×2) reconstruction have been appeared (Fig. 2). A very detailed electronic structure analyse of this clean surface can also be seen in ref. [48,52].

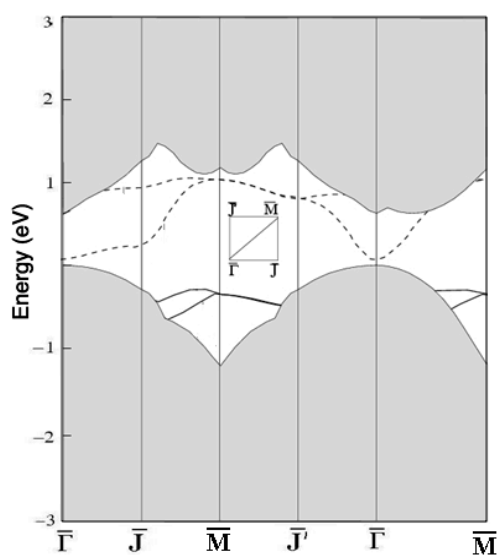

**Figure 2.** Energy band diagram of clean Si(001)-(2×2) surface

**3.2 Atomic Structure of Thiadiazole/Si(001)-(2×2) Surface**

In this study, both [4+2] and [2+2] cycloaddition mechanisms and bridge-type bonding models for adsorption of each of the four thiadiazole isomers on the Si(001) surface have been considered. As seen from Fig.3, only one [4+2] cycloaddition model and two bridge-type bonding models have been considered for all of the isomers. As for [2+2] cycloaddition mechanism, two models for 1,2,3- and 1,2,4-Thiadiazole have been offered depending on the position of the heteroatoms in the structure, but only one possible model is allowed for 1,2,5- and 1,3,4-Thiadiazole adsorption. The lattice geometries and ionic positions of the considered models have been relaxed firstly in order to determine the most stable ones among each isomer group. The calculated total and adsorption energies can be seen in Table 1. Total energy values are those obtained as the result of relaxation

process. However, the adsorption energies were calculated according to the formula of

$$E_{Adsorption} = E_{Si+Thiadiazole} - (E_{Si} + E_{Thiadiazole})$$

**Table 1.** Calculated relaxation and adsorption energies of the surfaces. $E_{Rel}$ is used for relaxation energy, while $E_{Ad}$ is used for adsorption energy. All energy values are given in eV.

|  | Total Energy of Single Molecule | [2+2] - I | | [2+2] - II | | [4+2] | | Bridge - I | | Bridge - II | |
|---|---|---|---|---|---|---|---|---|---|---|---|
|  |  | $E_{Rel}$ | $E_{Ad}$ | $E_{Rel}$ | $E_{Ad}$ | $E_{Rel}$ | $E_{Ad}$ | $E_{Rel}$ | $E_{Ad}$ | $E_{Rel}$ | $E_{Ad}$ |
| 1,2,3- | 0 | 0 | -1.16 | -0.12 | -1.28 | -0.13 | -1.29 | -0.46 | -1.62 | -0.26 | -1.42 |
| 1,2,4- | -0.73 | 0 | -0.93 | -0.36 | -1.29 | -0.28 | -1.22 | -0.64 | -1.58 | -0.56 | -1.49 |
| 1,2,5- | -0.60 | -0.10 | -1.03 | - | - | 0 | -0.93 | -0.53 | -1.46 | -0.72 | -1.65 |
| 1,3,4- | -0.06 | -0.34 | -1.40 | - | - | -0.25 | -1.31 | -0.50 | -1.56 | 0 | -1.06 |

As seen from the Table 1, energetically most favourable ones among all considered models for each of the isomers are bridge-type bonding models. When the total energy values of each isomer group are compared within themselves, it is seen that the Bridge-I model for 1,2,3-, 1,2,4- and 1,3,4-Thiadiazole have the lowest energy whereas Bridge-II for 1,2,5-Thiadiazole is more stable. This is actually an expected result because the dimer components of the clean silicon surface (which are electronically more unstable than the bulk due to the differentiation of the crystal structure on the surface) are saturated with the adsorption of molecules. During the adsorption process, silicon dimers become symmetric due to the charge transfer from electron-rich upper component to the electron-deficient lower component. Thus, the π electrons in the double bond of molecule can react with the π bond of the symmetric Si dimer. However, in [2 + 2] and [4 + 2] cycloaddition reactions, one of the dimers remains asymmetric with a tilt angle and this empty dimer may lead to a higher energy due to the unsaturated electronic configuration. This result is consistent with refs [56] and [57], in which the bridge-type adsorption of thiophene (a 5-membered aromatic ring with the formula of $C_4H_4S$) and benzene (a 6-membered aromatic ring with the formula of $C_6H_6$) on Si(001) was found to be more stable than [4+2] cycloaddition, respectively.

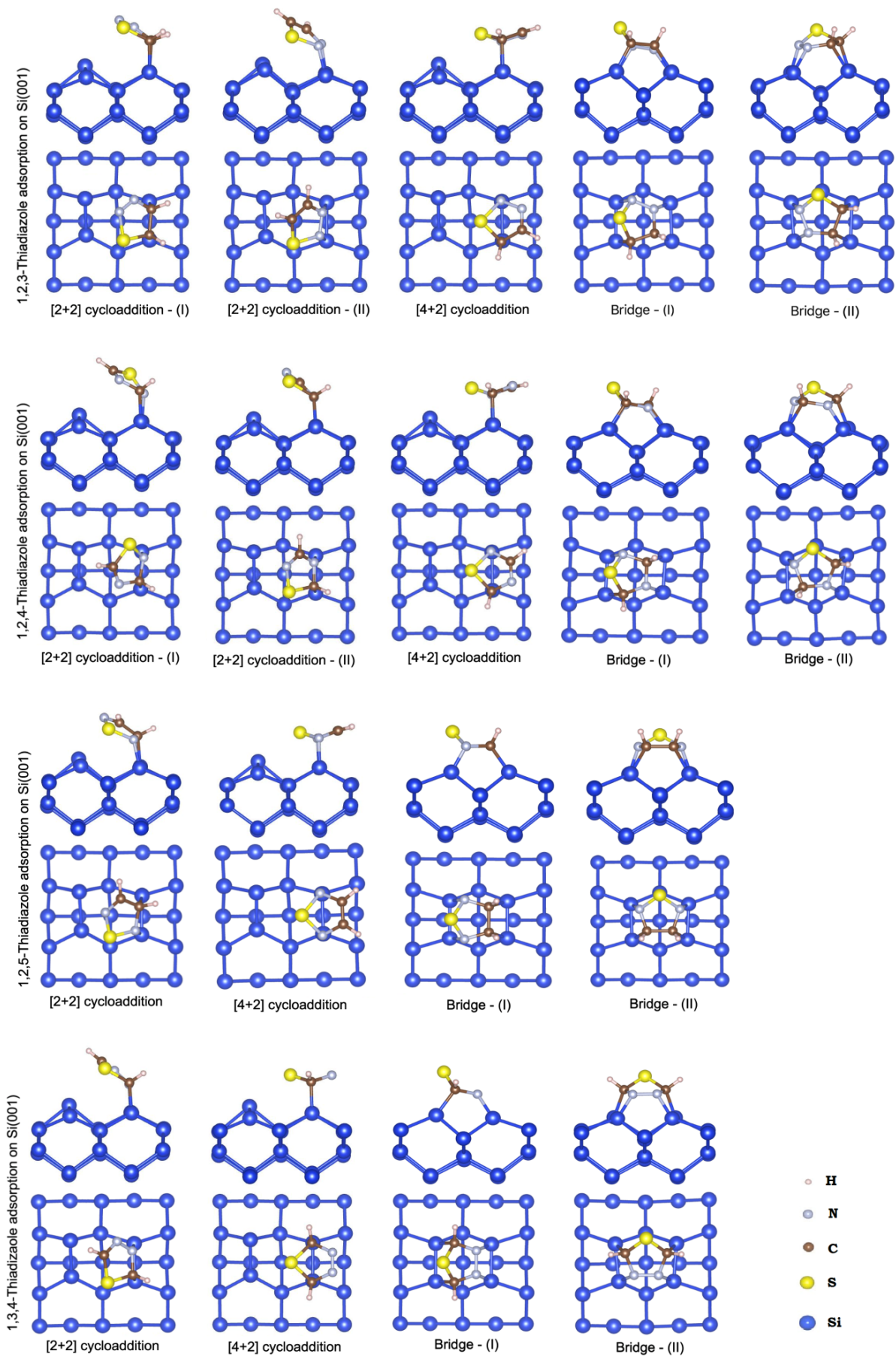

**Figure 3.** Top and side views of considered models

The negativity of the adsorption energy values in Table 1 indicates that the studied models are chemically stable and these reactions are exothermic. This result is in agreement with ref [58], which studied [4+2] cycloaddition reactions of various aromatic ring compounds on the Si(001)-(2x1) surface and showed that all these reactions are exothermic. As can be seen from Table 1, it is obvious that in an isomer group, the absolute value of the adsorption energy of the most stable model (as determined by $E_{Rel}$) will be greater than other models of the same isomer. In this case, when the adsorption energies of the four most stable structures of four different isomers are compared among themselves, it is seen that the Bridge-II type bonding of 1,2,5-Thiadiazole has a slightly larger value, although the single 1,2,5-Thiadiazole molecule is not the most stable one among all isomers.

The calculated atomic key parameters for the most stable surfaces are given in Table 2. Since there is no data on thiadiazole adsorption on silicon in the literature, experimental and theoretical results of single molecules are given as the references in Table 2. According to the table, almost all the bond lengths of the adsorbed molecule are seen to be elongated compared to the free molecule. Also the angle values given in the references [59-61] can be regarded as compatible with those calculated here. Small differences are due to the fact that the molecule loses its planar structure after adsorption process. Bond lengths of the silicon dimers are slightly increased when compared to the clean surface. But these values are still inside the range of 2.20–2.47Å obtained experimentally for the Si(001). Thus, the Si–Si dimer is not broken during the adsorption of thiadiazole. But, this increase in bond lengths indicates that the strength of Si-Si bonds is weakened due to molecular adsorption or in other words due to the changes in electronic nature. Also, the calculated Si-Si dimer, Si-C and C-C bond lengths are matching with the values previously found for furan [24] and benzene [62] adsorption.

**Table 2.** The calculated atomic key parameters of the most stable surfaces. (Bond lengths are in Å. †, + and * refer to the values given in ref. [59], [60] and [61], respectively. The indices (1) and (2) represent the left and right dimes, respectively, in Fig. 3.)

| | | Bridge - I 1,2,3- Thiadiazole | Bridge - I 1,2,4-Thiadiazole | Bridge - II 1,2,5-Thiadiazole | Bridge - I 1,3,4-Thiadiazole |
|---|---|---|---|---|---|
| **Bond Lengths** | Si-Si (1) | 2.38 | 2.39 | 2.34 | 2.38 |
| | Si-Si (2) | 2.33 | 2.33 | 2.34 | 2.33 |
| | Si-C (1) | 1.96 | 2.0 | 2.0 | 2.03 |
| | Si-C (2) | 2.03 | 2.0 | 2.0 | 2.03 |
| | Si-N (1) | 1.88 | 1.84 | 1.81 | 1.87 |
| | Si-N (2) | 1.82 | 1.82 | 1.81 | 1.87 |
| | C-C | 1.55 1.37† | - | 1.61 1.44* | - |
| | N-N | 1.47 1.29† | - | - | 1.50 1.36* |
| | N-C | 1.50 1.37† | 1.47/1.49/1.52 1.31/1.32/1.37+ | 1.50 1.28* | 1.46 1.27* |
| | S-N | 1.72 1.69† | 1.74 1.65+ | 1.72 1.64* | - |
| | S-C | 1.86 1.69† | 1.85 1.71+ | - | 1.83 1.73* |
| | C-H | 1.10 1.08† | 1.10 1.08+ | 1.10 1.07* | 1.10 1.07* |
| **Angles** | | (1,2,3-thiadiazole: 105°, 90°, 112°, 107°, 98°) | (1,2,4-thiadiazole: 101°, 89°, 107.5°, 110.5°, 103°) | (1,2,5-thiadiazole: 94°, 106.5°, 106.5°, 107.5°, 107.5°) | (1,3,4-thiadiazole: 104°, 85°, 109.5°, 100.5°, 104°) |

### 3.3 Electronic Structure of Thiadiazole/Si(001)-(2x2) Surface

In this part of the study, electronic band structure of the four stable Thiadiazole/Si(001)-(2x2) surfaces and their corresponding total density of states (TDOS) have been plotted to see how the molecule adsorption affects the electronic structure of the clean surface. As shown in Fig. 4, electronic band graphics of the thiadiazole surfaces have been superimposed over the projected bands of bulk Si(001)-(2x2) to determine the new surface states appear in the fundamental gap of bulk silicon due to the adsorption. In addition, partial charge densities of these surface states have been calculated to show the origin of them. In this method, charge density of each individual energy band have been decomposed according to each point of the k-mesh, which helps to understand which atoms and orbitals bring out the surface states. Here, partial charge densities have been plotted at M-point of the k-mesh. The TDOS graphics have been obtained by superimposing the thiadiazole surfaces on the dimerized clean silicon surface to see the changes in the electronic structure of the substrate.

All of the TDOS plots show that the clean silicon surface is passivated and the band gap is widened by the adsorption of the molecule. But this change is slightly less in 1,3,4-Thiadiazole than in the others. Similarly, the energy band graphics of these isomers show that the empty surface states of the clean surface are totally pushed out of the main band gap region.

There are two occupied surface states ($S_1$ and $S_2$) in 1,2,3- and 1,2,5-Thiadiazole adsorption, and three occupied surface states ($S_1$, $S_2$ and $S_3$) in 1,2,4- and 1,3,4-Thiadiazole adsorption. They are all located below the Fermi level, which causes the corresponding surfaces to have a semiconductor character. 1,2,3-Thiadiazole/Si(001)-(2×2) and 1,2,4-Thiadiazole/Si(001)-(2×2) surfaces have an indirect band gap of 0.848 eV and 0.933 eV around the Γ point, respectively. 1,2,5-Thiadiazole/Si(001)-(2×2) surface has an indirect band gap of 0.856 eV from the maximum of the valence band at between the J-M points to the minimum of the conductivity band at the Γ point. 1,3,4-Thiadiazole/Si(001)-(2×2) surface has an indirect band gap of 0.638 eV from M to Γ point, which is larger than the dimerized clean surface but almost same with the bulk silicon. The surface states are mostly originated from the p-orbitals of the molecule components, but contribution from S and N atoms is more dominant than the C atoms in 1,2,3-, 1,2,4- and 1,2,5-Thiadiazole. As to 1,3,4-Thiadiazole, only the N atom is dominant in $S_1$, while the greatest contribution to $S_2$ and $S_3$ is made by sulfur.

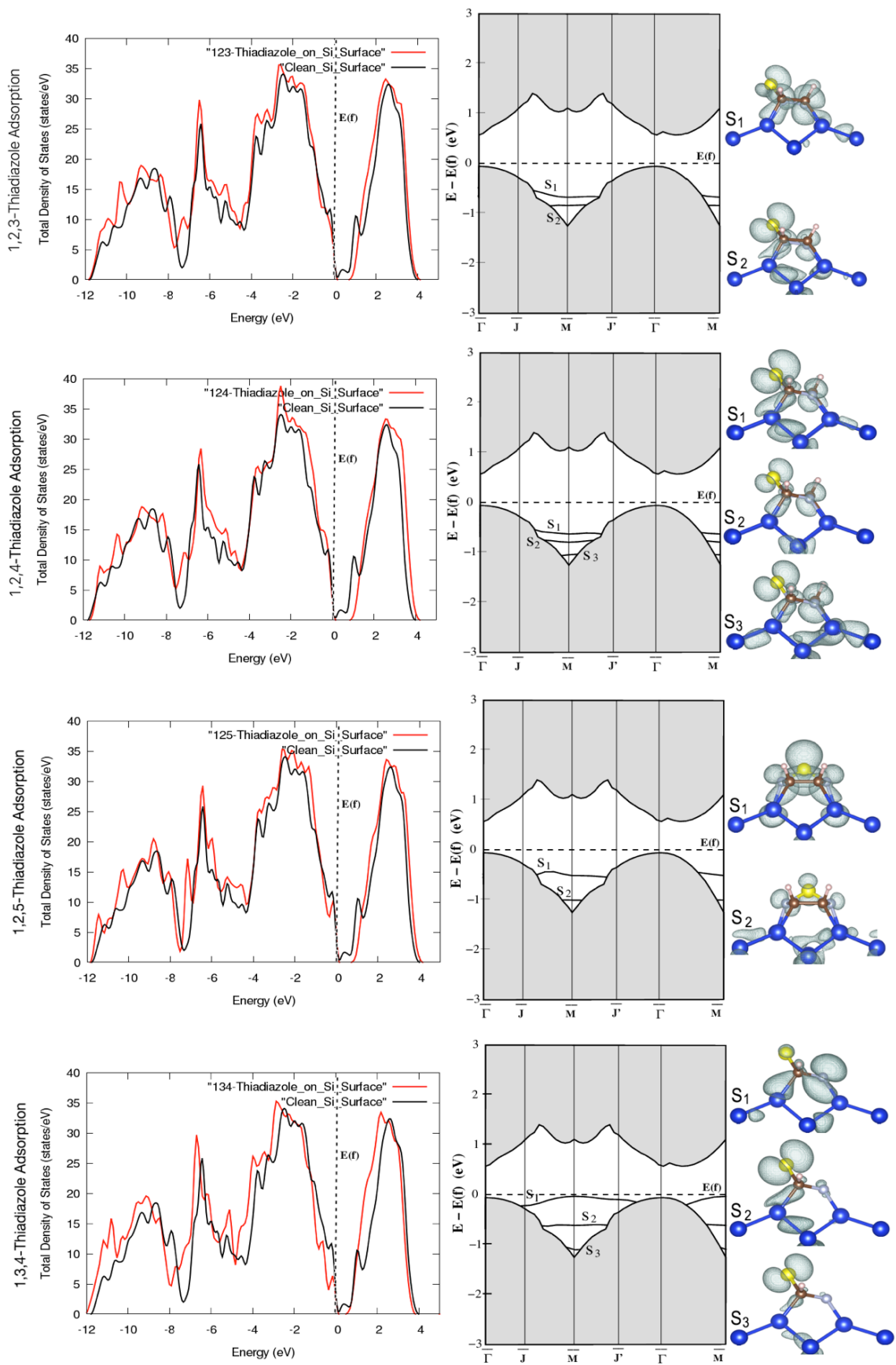

**Figure 4.** Total Density of States, Energy Band graphics and corresponding charge densities of the Thiadiazole/ Si(001)-(2×2) surfaces.

The PDOS graphics of the stable structures have been plotted in Fig. 5 to see the partial contributions of s- and p-orbitals over the entire DOS. As seen from the figures, p-orbitals exhibit a more dominant character around the band gap region, while s-orbitals become more visible as they move away from the gap region. These results are consistent with the plots in Fig. 4 showing that the charge densities of surface states around the band gap are mostly due to p-orbitals.

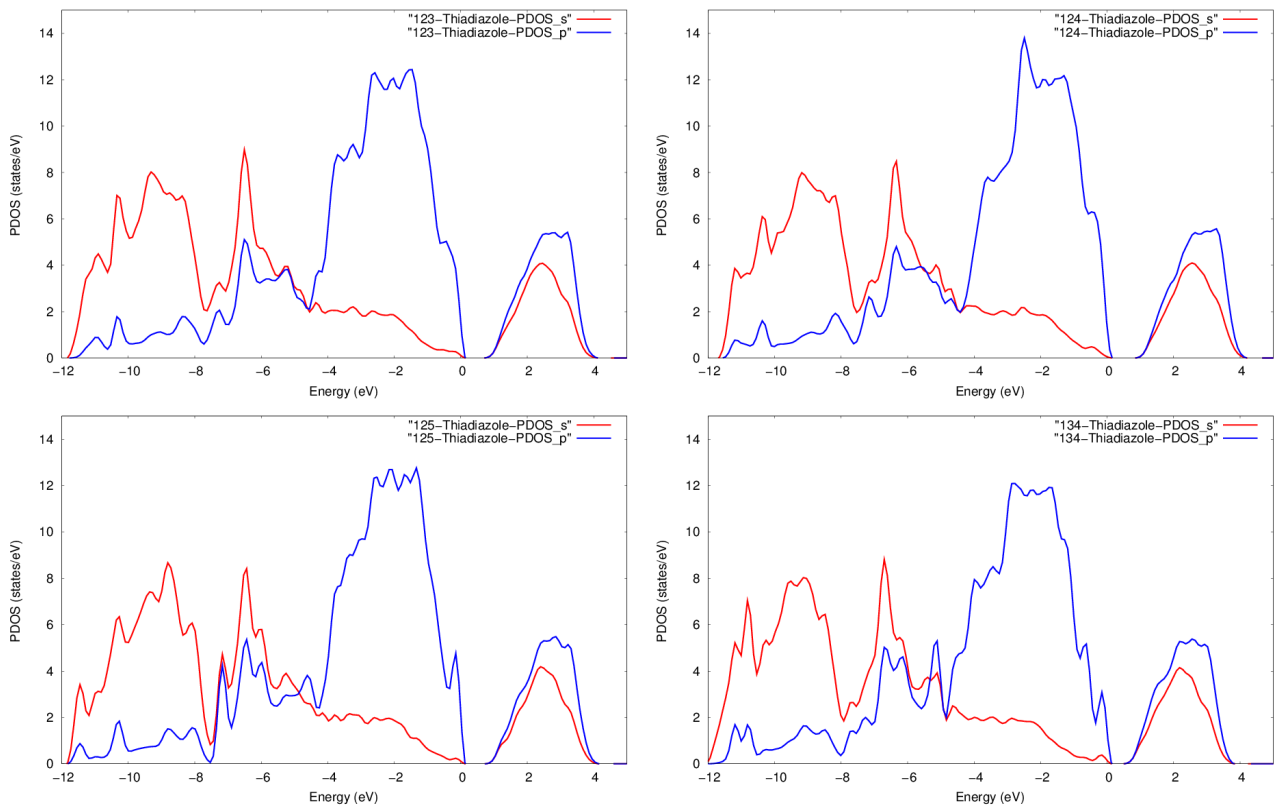

**Figure 5.** PDOS graphics of the Thiadiazole/Si(001)-(2×2) surfaces.

In the last section of study, charge density differences induced by thiadiazole adsorption on silicon surface have been calculated with the formula of $\Delta\rho = \rho_{Si+Thiadiazole} - (\rho_{Si} + \rho_{Thiadiazole})$, and plotted in Fig. 6. Here, $\Delta\rho = \rho_{Si+Thiadiazole}$ represents the charge density of the thiadiazole adsorbed structure, while $\rho_{Si}$ and $\rho_{Thiadiazole}$ are the charge densities of identical structures. The isosurface value is ±0.01 e$^-$Å$^{-3}$ and red (blue) colour symbolizes the electron gain (loss). In Fig. 6, charge depletion areas in between N and C atoms indicate that double bonds in the molecule are broken, which is consistent with the increase in bond distances in Table 2. However, a large charge accumulation between molecule and substrate is clearly seen due to the new bond formation.

Electronegativity of the elements in the structure can provide an extra information to decide the direction of charge transfer. On Pauling scale, the electronegativity of Si is 1.90, while it is 2.55, 3.04 and 2.58 for C, N and S, respectively. Thus the direction of the charge transfer can be expected to be from the substrate to the molecule, which supports the existence of red areas mostly on the molecule.

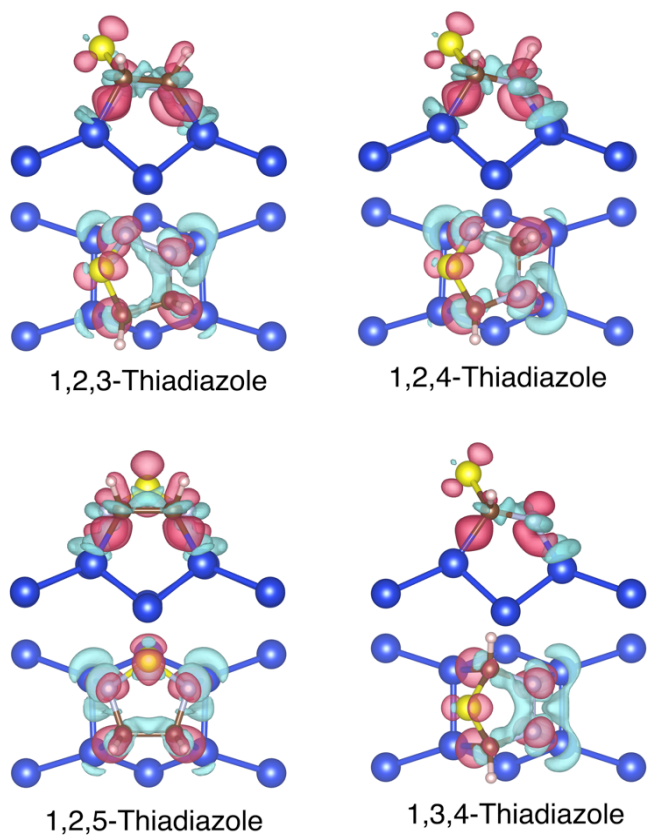

**Figure 6.** Charge density difference isosurfaces. The isosurface value is $\pm 0.01$ e$^-$Å$^{-3}$. Red (blue) colour symbolizes the electron gain (loss).

## 4. Conclusion

The effect of thiadiazole adsorption on the structural and electronic properties of clean Si(001)-(2×2) surface have been studied by using DFT methods. All of the four structural isomers of this molecule have been considered as the adsorbate. A total of eighteen bonding mechanisms have been simulated. Bridge-type bonding models have been found more favourable than [2+2] and [4+2] cycloaddition. Among all considered structures, 1,2,5-Thiadiazole has the biggest adsorption energy of 1.65 eV. When the magnitude of the adsorption energies and charge density differences are discussed together, thiadiazole adsorption on the silicon surface can be evaluated as a chemisorption. According to the electronic band diagrams of the stable structures, 1,2,4-Thiadiazole adsorption has the largest band gap of 0.933 eV and 1,3,4-Thiadiazole has the minimum band gap of 0.638 eV. Energy band gap of 1,2,3- and 1,2,5-Thiadiazole have a similar value; 0.848 eV and 0.856 eV, respectively. By comparing Fig. 2 and Fig. 4, it is seen that the band gap of the clean surface is widened due to the adsorption of thiadiazole which may be useful for some tailored technological applications. This result is consistent with other studies showing that the cyclic organic compounds passivate the silicon surface [24,63].

**Acknowledgement**

The study is supported by Ministry of Development of Turkey under Grant No: DPTK2006K120470.

**References**

[1] S. Srivastava, R. K. Prasad, R. Saini, WJPPS, 3 (2014) 1198.

[2] D. Glossman-Mitnik, Journal of Molecular Structure: Theochem 549 (2001) 285.

[3] A. T. Balaban, D. C. Oniciu, A. R. Katritzky, Chem. Rev. 104 (2004) 2777.

[4] M.D. Glossman J. Mol. Struct. Theochem 330 (1995) 385.

[5] F. Hipler, S. Gil Girol, W. Azzam, R. A. Fischer, C. Wöll, Surface Langmuir 19 (2003) 6072.

[6] Y. Hu, C. Y. Li, X. M. Wang, Y. H. Yang, H. L. Zhu, Chem. Rev. 114 (2014) 5572.

[7] J.Dwivedi, N.Kaur, D. Kishore, S. Kumari and S.Sharma, Current Topics in Medicinal Chemistry, 16 (2016) 2884.

[8] K. Chattrairat, D. Phromyothin, Materials Today: Proceedings 4 (2017) 6118.

[9] L. Joseph,, M. George, P. Mathews, J Pharm Chem Biol Sci. 3(3) (2015) 329.

[10] Y. Li, J. Geng, Y. Liu, S. Yu, G. Zhao, ChemMedChem 8 (2013) 27.

[11] K. Ajay Kumar, G. Vasanth Kumar, N. Renuka, Int. J. Pharm Tech. Res. 5(1) (2013) 239.

[12] R. Solmaz, G. Kardas, B. Yazici, M. Erbil, Colloid Surf. A 312 (2008) 7.

[13] J. Zhou, S.Chen, L.Zhang, Y. Feng, H. Zhai, J. Electroanal. Chem. 612 (2008) 257.

[14] M. Sato, T. Kamita, K. Nakadera, and K.-I. Mukaida, Eur. Polym. J., 1995, 31, 395.

[15] P. A. Koutentis, C. P. Constantinides, Comprehensive Heterocyclic Chemistry III 5 (2008) 566.

[16] R. T. Loto, C. A. Loto, A.P.I. Popoola, J. Mater. Environ. Sci. 3(5) (2012) 885.

[17] W. Xue, W. Ma, X. Xu, T. Li, X. Zhou, P. Wang, Industrial Lubrication and Tribology 69 (2017) 891.

[18] Z. Fan, Z.Shi, H. Zhang, X. Liu[†], L. Bao, L. Ma, X. Zuo, Q. Zheng and N. Mi, J. Agric. Food Chem., 57 (2009) 4279.

[19] A. Vainer, K. M. Dyumaev, S. K. Lodygin, N. P. Posadskaya, Doklady Chemistry 383 (2002) 120.

[20] W. Dehaen, V. A. Bakulev, E. C. Taylor, J. A. Ellman, The Chemistry of Heterocyclic Compounds, The Chemistry of 1, 2, 3-Thiadiazoles, John Wiley & Sons, 2004.

[21] M. Mamada, H. Shima, Y. Yoneda, T. Shimano, N. Yamada, K. Kakita, T. Machida, Y. Tanaka, S. Aotsuka, D. Kumaki, S. Tokito, Chem. Mater. 27 (2015) 141.

[22] L. Wang, B. Dong, R. Ge, F. Jiang, J. Xiong, Y. Gao, J. Xu, Microporous Mesoporous Mater. 224 (2016) 95.

[23] Y. Jin, Z.Chen, M. Xiao, J. Peng, B. Fan, L.Ying et. al., Adv. Energy Mater. 7 (2017) 1700944.

[24] Ç. Kaderoglu, B. Kutlu, B. Alkan, M. Çakmak, Surface Science 602 (2008) 2845.

[25] Z. Hossain, T. Aruga, N. Takagi, T. Tsuno, N. Fujimori, T. Ando, M. Nishijima, Jpn. J. Appl. Phys. 38 (1999) L 1496.

[26] S. F. Bent, Surface Science 500 (2002) 879.

[27] G. Ashkenasy, D. Cahen, R. Cohen, A. Shanzer, A. Vilan, Acc. Chem. Res. 35 (2002) 121.

[28] S. Ohno, H. Tanaka, K. Tanaka, K. Takahashi, M. Tanaka, Organic Electronics 25 (2015) 170.

[29] S. Godlewski, H. Kawai, M. Engelund, M. Kolmer, R. Zuzak, A. Garcia-Lekue, G. Novell-Leruth, A. M. Echavarren, D. Sanchez-Portal, C. Joachim, M. Saeys, Phys.Chem.Chem.Phys. 18 (2016) 16757.

[30] F. Gao, A. V. Teplyakov, Applied Surface Science 399 (2017) 375.

[31] E. Seo, D. Eom, J-M. Hyun, H. Kim, J-Y. Koo, Surface Science 656 (2017) 33.

[32] H. G. Huang, J. Y. Huang, Y. S. Ning, G. Q. Xu, J. Chem. Phys., 121 (2004) 4820.

[33] F. Tao, S. L. Bernasek, J. Am. Chem. Soc. 129 (2007) 4815.

[34] Y. Cao, K. S. Yong, Z. H. Wang, J. F. Deng, Y. H. Lai, G. Q. Xu, J. Chem. Phys. 115 (2001) 3287.

[35] G. Kresse, J. Hafner, Phys. Rev. B 47 (1993) 558.

[36] G. Kresse, J. Hafner, Phys. Rev. B 49 (1994) 14251.

[37] G. Kresse, J. Furthmuller, Comp. Mat. Sci. 6 (1996) 15.

[38] G. Kresse, J. Furthmuller, Phys. Rev. B 54 (1996) 11169.

[39] P.E. Blochl, Phys. Rev. B 50 (1994) 17953.

[40] G. Kresse, D. Joubert, Phys. Rev. B 59 (1999) 1758.

[41] J. P. Perdew, K. Burke, and M. Ernzerhof, Phys. Rev. Lett. 77 (1996) 3865.

[42] J. P. Perdew, J. A. Chevary, S. H. Vosko, K. A. Jackson, M. R. Pederson, D. J. Singh, and C. Fiolhais, Phys. Rev. B 46 (1992) 6671.

[43] H. J. Monkhorst, and J. D. Pack, Phys. Rev. B 13 (1976), pp. 5188.

[44] W. H. Press, B. P. Flannery, S. A. Teukolsky, and W. T. Vetterling, Numerical Recipes (Cambridge University Press, New York, 1986).

[45] K. Momma, F. Izumi, J. Appl. Crystallogr. 44 (2011) 1272.

[46] D.J. Chadi, Phys. Rev. Lett. 43 (1979) 43.

[47] P. Krüger, J. Pollmann, Phys. Rev. Lett. 74 (1995) 1155.

[48] J. Fritsch, P. Pavone, Surf. Sci. 344 (1995) 159.

[49] A.W. Munz, C.H. Ziegler, W. Göpel, Phys. Rev. Lett. 74 (1995) 2244.

[50] D. Badt, H. Wengelnik, H. Neddermeyer, J. Vac. Sci. Technol. B 12 (1994) 2015.

[51] S.C.A. Gay, G.P. Srivastava, Phys. Rev. B 60 (1999) 1488.

[52] M. Cakmak, E. Mete, S. Ellialtioglu, Surface Science 600 (2006) 3614.

[53] P. Becker, P. Seyfried, and H. Siegert, Z. Phys. B - Condensed Matter 48 (1982) 17.

[54] Çakmak, M., Theoretical studies of structural and electronic properties of overlayer on semiconductor surfaces. Doctoral Thesis, University of Exeter, Dept. of Physics, Exeter, 1999.

[55] Ç. Kaderoglu, B. Alkan, M. Çakmak, Surface Science 606 (2012) 470.

[56] N. Isobe, T. Shibayama, Y.Mori, K. Shobatake, K. Sawabe, Chem. Phys. Lett. 443 (2007) 347.

[57] J-Y. Lee and J-H. Cho, Phys. Rev. B 72 (2005) 235317.

[58] X. Lu, M. C. Lin, X. Xu, N. Wang and Q. Zhang, Phys. Chem. Comm. 13 (2001) 1.

[59] O.L. Stiefvater, Chemical Physics, 13 (1976) 73.

[60] O.L. Stiefvater, Z. Naturforsch. A 31 (1976) 1681.

[61] M.D. Glossman J. Mol. Struct. Theochem 330 (1995) 385.

[62] Y. Jung and M. S. Gordon, J. Am. Chem. Soc. 127 (2005) 3131.

[63] S. Y. Quek, J. B. Neaton, M. S. Hybertsen, E. Kaxiras, and S. G. Louie, Phys. Stat. Sol. B, 243, (2006) 2048.